\pdfoutput=1

\documentclass{iaus}
\usepackage{graphicx}

\newcommand{\halpha}{H$\alpha$}
\newcommand{\mi}{$\mu$m}

\title[Dust in dwarf galaxies] 
{Dust properties and distribution in dwarf galaxies}

\author[Ute Lisenfeld et al.]   
{Ute Lisenfeld$^{1,2}$,
Monica Rela\~no$^{1,2}$,
Jos\'e V\'\i lchez$^3$,
 Eduardo Battaner$^1$
 \and Israel Hermelo$^1$}

\affiliation{$^1$Universidad Granada, Spain\\ email: {\tt ute@ugr.es, battaner@ugr.es} \\[\affilskip]
$^2$Institute of Astronomy, University of Cambridge, UK  \\email: {\tt mrelano@ast.cam.ac.uk} \\[\affilskip]
$^3$Instituto de Astrof\'isica de Andaluc\'ia, Granada, Spain  \\email: {\tt jvm@iaa.es}}

\pubyear{2008}
\volume{255}  
\pagerange{119--126}
\jname{Low-Metallicity Star Formation: From the First Stars to Dwarf Galaxies}
\editors{L.K. Hunt, S. Madden \& R. Schneider, eds.}
\begin{document}

\maketitle

\begin{abstract}
We present a study of the extinction, traced by the Balmer decrement, in HII regions
in the dwarf galaxies NGC 1569 and NGC 4214. 
We find that the large-scale extinction around the most prominent HII regions in both
galaxies forms a shell in which locally the intrinsic extinction can adopt
relatively high values ($A_V = 0.8 - 0.9$ mag) despite the low metallicity and thus
the low overall dust content. 
The small-scale extinction (spatial resolution $\sim$0.3'') shows fluctuations
that are most likely due to variations in the dust distribution.
We compare the distribution of the
extinction to that of the dust emission, traced by {\it Spitzer} emission at  8 and 24\mi,
and to the emission of cold dust at 850\mi.
We find in general a good agreement between all tracers, expect for the 850\mi\
emission in NGC 4214 which is more extended than the extinction
and the other emissions.
Whereas in NGC 1569 the dust emission at all wavelengths
is very similar, NGC 4214 shows spatial variations in the  24-to-850\mi\ ratio.

We furthermore compared the 24\mi\  and the extinction-corrected
\halpha\ emission from HII regions in a sample of galaxies with a wide range of metallicities
and found a good correlation between both emissions, independent of metallicity. 
We suggest that  this
lack of dependence on metallicity might be due to the formation of dust shells with a 
relatively constant opacity, like
the ones observed here,  around ionizing stars.

\end{abstract}

\firstsection 
\section{Introduction}

Interstellar dust can be studied via its emission and also via the extinction that 
it causes. Each method has its advantages and difficulties.
The most common way of obtaining  the extinction  of the light coming from
HII regions is based on the comparison of the \halpha\ and H$\beta$ recombination
line fluxes.
A major advantage of this method is the high spatial resolution achieved,
determined by the resolution of the optical images. It is, however, difficult
to derive the distribution of the dust from extinction maps 
because the relative distribution of the dust and the gas plays a mayor role and because
maps of the \halpha/H$\beta$ ratio are biased towards low-extinction regions.

Alternatively, the dust can be studied via its emission,
which depends on the dust amount, the type of grains and the dust temperature
determined by the interstellar radiation field (ISRF).
Although very different models for the interstellar dust exist (see, e.g., Zubko et al. 2004 and
references therein), they 
 generally need to include three types of grains: big grains that are
in equilibrium with the ISRF, very small grains (VSGs) that are stochastically
heated and are necessary to explain the mid-infrared emission 
and Polycyclic Aromatic Hydrocarbons (PAHs) to explain the line features 
in the 8 \mi\ range.
So far there have been two major problems for studies of the
dust emission: The lack of images at a high spatial resolution and
the lack of submillimeter data probing the cold dust. Fortunately, the situation is
improving.
The  {\it Spitzer} satellite has for the first time provided images of
high spatial resolution in the mid-infrared regime where the
emission is dominated by PAHs and VSGs.
In the near future, the satellites {\it Herschel} and {\it Planck} will
provide a large database for the submillimeter emission of galaxies.

We present a study of  the dust extinction (based on the Balmer decrement)
and emission in two
nearby, starbursting, dwarf galaxies,  NGC 1569 and NGC 4214.
The goals are  to derive the small and large-scale structure of the
dust  extinction and to compare it to the dust emission at various wavelengths,
tracing different types of dust grains and temperatures.
This study will allow us to derive information about the dust distribution
and variations of the dust temperature and of grain types.

NGC 1569 and NGC 4214 have been chosen in order to study of 
role of the metallicity and active star formation for  the dust.
Both galaxies have a low metallicity  of 12+log(O/H) = 8.2 (Kobulnicky \& Skillman 1997, 
Armus et al. 1989) and a similar distance
(2.2 Mpc to NGC 1569,  Israel 1988,
and 2.9 Mpc to NGC 4214, Ma\'iz-Apell\'aniz et al. 2002).
They both harbour large HII complexes, and host 
one (NGC 4214) or two (NGC 1569) Super Star Clusters (SSC) in the middle
of a large expanding \halpha\ shell (Martin 1998).

\section{Extinction maps}

The extinction maps (A$_{H\alpha}$) were calculated using \halpha\ and H$\beta$ 
images obtained with the
WFPC2 on the Hubble Space Telescope and
following the method explained in Caplan \& Deharveng (1986). 
In NGC 1569, an extinction map of the galaxy was obtained in two different regimes
characterized by different physical conditions: extinction within
the HII regions and extinction of the diffuse ionized gas (see Rela\~no et al 2006, for
a detailed explanation of the procedure). In NGC 4214, the extinction was only
calculated for the HII regions and the diffuse ionized gas was excluded.
In both galaxies, we furthermore excluded regions with  a low H$\beta$ equivalent width 
(values below 75 \AA)  because of the possible stellar absorption of 
H$\beta$ due to  evolved stars.

\subsection{Large-scale distribution}

In Fig. 1 we show the resulting extinction maps overlaid with
the \halpha\ emission distribution, both smoothed to a resolution
of 6'' in order to make the large-scale distribution visible.
We see various interesting features that are in common in both galaxies:

\begin{itemize}
\item The highest values in extinction form a shell structure 
around the brightest HII region.
The shells are also visible in \halpha\  but much clearer in the extinction.
They spatially correlate with expanding supershells catalogued by 
Martin (1998) (NGC 1569-C, respectively NGC 4214-A).

\item The values of the intrinsic extinction (i.e. after subtracting the foreground
Galactic extinction) in the shells are A$_{H\alpha}$ =  0.3 - 0.7 mag in both galaxies.
These value are relatively high for these low-metallicity galaxies,
indicating that locally
high-opacity regions can exist, in spite of the globally low dust content.

\end{itemize}

We suggest that the shell-like extinction structure has been formed by a cumulative deposit
of dust at the boundary of the supershell. In NGC 1569, an analysis of the
energies involved shows that the
stellar winds coming from SSC A are able to produce the expanding shell and to sweep up
the dust mass contained in it (Rela\~no et al. 2006).

\begin{figure}[b]
\begin{center}
\centerline{
\includegraphics[width=6.cm,angle=270]{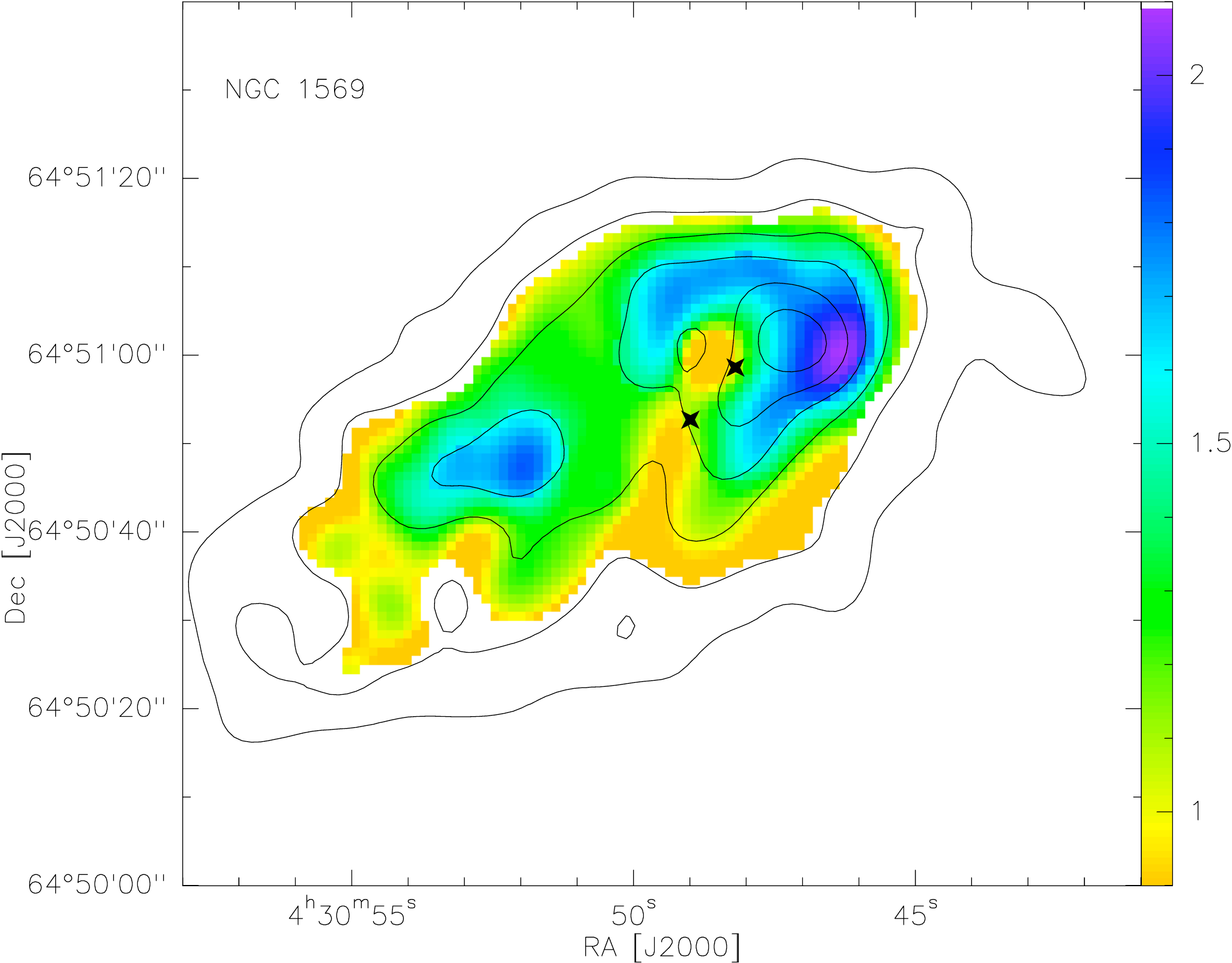} \quad
\includegraphics[width=6.cm,angle=270]{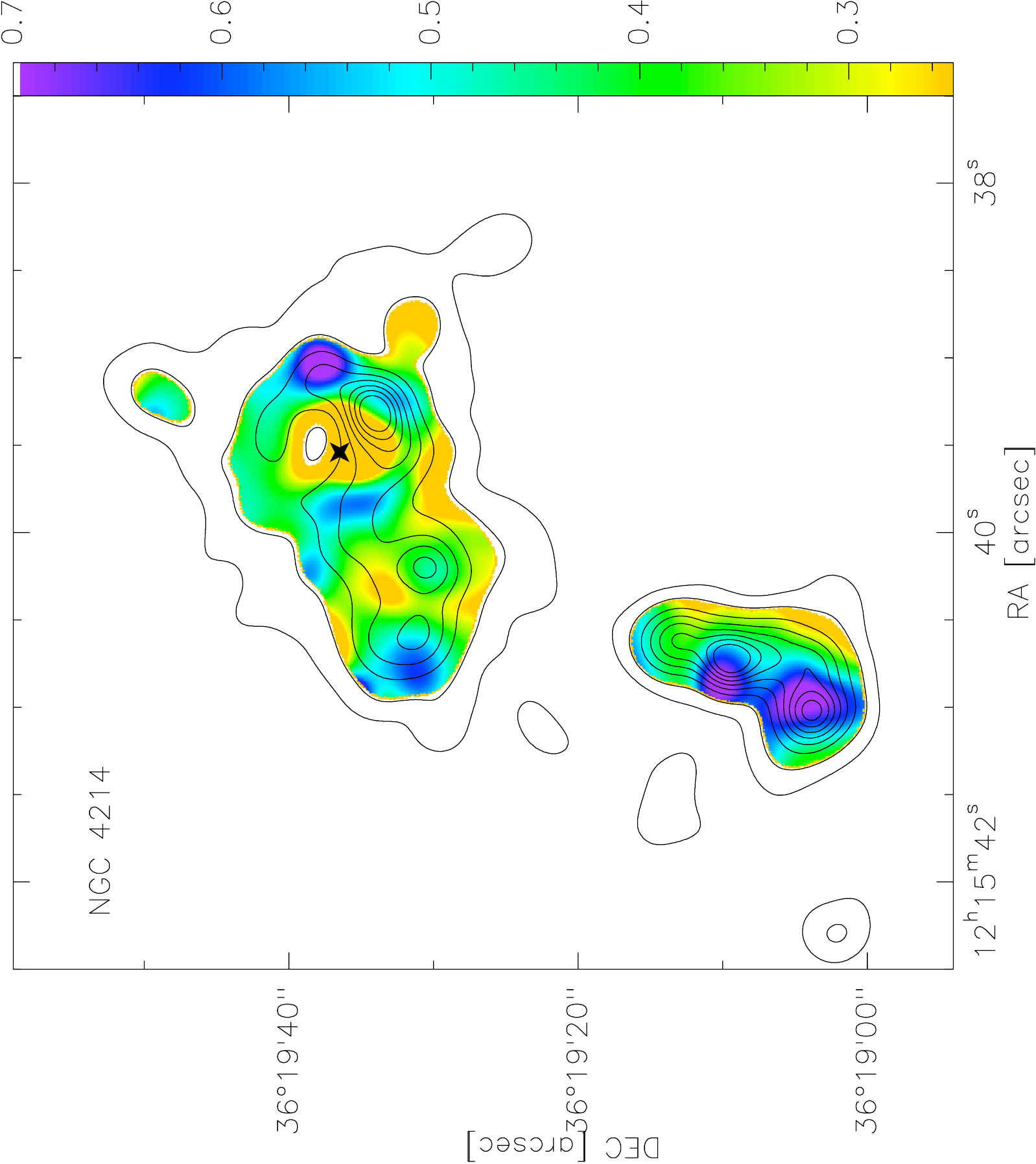} 
  }
 \caption{Total (i.e. intrinsic plus Galactic foreground)
 extinction, $A_{H\alpha}$, (color),  overlaid on the emission of  \halpha\  (contours) for
 ({\it left}) NGC 1569 and ({\it right}) NGC 4214.The resolution of both images is 6''. 
The Galactic foreground extinction is about $A_{H\alpha} = 1.3$ mag for NGC 1569 (Kobulnicky \& Skillman 1997)
and $A_{H\alpha} = 0.05$ mag for NGC 4214 (Schlegel et al. 1998). The crosses indicate
the positions of the SSCs A and B in NGC 1569, and SSC I-As in NGC 4214.
}
\end{center}
\end{figure}

\subsection{Small scale distribution}

In the 0.3'' resolution \halpha/H$\beta$ image a very clumpy structure
is visible (see Fig. 3 of Rela\~no et al. 2007 for NGC 1569). 
In both galaxies,  the fluctuations are clearly above the level expected 
from the noise in the \halpha\ and H$\beta$ images.
We suggest that these fluctuations in  \halpha/H$\beta$ can be
related to opacity inhomogeneities in the HII regions, most likely
due to a clumpy dust distribution.

\section{Comparison of the dust extinction and emission}

In Fig. 2 we show a comparison of the extinction and dust emission
at 8\mi, 24\mi\ and 850\mi.
We find a generally good, but not always perfect,  agreement between the
8\mi\ and 24\mi\ emission and the dust extinction.  There is a trend
for the 24\mi\  emission (in NGC 1569 also for the 8\mi\ emission) 
to be shifted towards the inner side of 
the extinction shell. This can be to a combination of two effects:
Towards the inner side of the HII region the dust is hotter and thus emits more
strongly at 24\mi. In addition, a geometric effect can explain the displacement
of the emission peaks: If the dust accumulates in the outer region of the ionized shell
a line of sight in this direction will trace a region of higher opacity than
further inside the HII region, where  more dust emission is seen due to the long line of
sight.

When comparing the dust extinction to the emission of the cold dust
at 850\mi\, observed with SCUBA at the James Clerk Maxwell
Telescope (Lisenfeld et al. 2001, Kiuchi et al. 2004), we find a different distribution in both objects:
In NGC 1569 the distribution at 850\mi\ is very similar to
that of \halpha, to the dust extinction, as well as to the dust emission at 8 and 24\mi.
Within the spatial resolution of 15'', the peak of the 850\mi\ emission agrees
with all emissions and with the dust extinction.
In NGC 4214, on the other hand, only part of the 850\mi\ emission 
coincides with the dust shell, and it is also  different from the dust emission
at 8\mi\ and 24\mi. 
In the southern HII region, there are two marginally resolved peaks
of the 850\mi\ emission 
close to peaks in the extinction and the \halpha\ emission.
In the north, part of the 850\mi\ emission spatially coincides with the 
extinction shell and the 24\mi\ emission, but the 850\mi\ emission extends
further northwest than any other emission considered here.
Thus, in contrary to NGC 1569, NGC 4214 shows very strong variations in the
24-to-850\mi\ ratio, indicating the presence of local variations of the
dust temperature or the grain composition, and in particular the presence
of cold dust in the northwestern part.

In both galaxies there is a close relation between the  molecular gas 
and the emission at 850\mi: In NGC 1569 giant molecular clouds are found
close to the peak of the 850\mi\ emission. 
In NGC 4214, on the contrary,  the correspondence is not complete (see Fig. 2).
In particular, towards the northwest no molecular gas
has been found
in spite of strong 850\mi\  emission, only atomic gas (Walter et al. 2001)
is present.


\section{The 24\mi\ emission as a SF tracer in low-metallicity galaxies}

The good agreement between the 24\mi\ and \halpha\ emission has
been found in several other studies in different kind of galaxies 
(Calzetti et al. 2005, 2007, P\'erez-Gonzalez et al. 2006) and it is the base for the
use of the 24\mi\ emission as a tracer for star formation (Calzetti et al. 2007,
Kennicutt et al. 2007).
Rela\~no et al. (2007) have shown by comparing  24\mi\ 
and extinction-correced \halpha\ emission integrated
over individual HII regions  in a sample of galaxies spanning a wide metallicity range
(from 12+log(O/H) = 7.2 to 9.1) that the correlation between both emissions does
not depend on metallicity.   Only when considering the total emissions
(i.e. the emission from HII regions plus the diffuse emission),
a trend  with metallicity was found.

This result is, at first sight,  surprising because
the 24\mi\ emission depends on the dust content which is directly 
related to the metallicity. 
The most likely reason of the observed constancy of the
\halpha/24\mi\ ratio of HII regions is the formation of a dust shell
like the ones observed in NGC 1569 and NGC 4214. 
In these shells, the opacity
might be relatively constant and not strongly dependent on the metallicity,  due
to  the existence of a lower threshold for the accumulation of
dust in HII regions in order to support large SFR 
( $\sim 10^{-4} - 10^{-1}$ M$_\odot$ yr$^{-1}$).

\begin{figure}[b]
\begin{center}
\centerline{\includegraphics[width=6cm,angle=270]{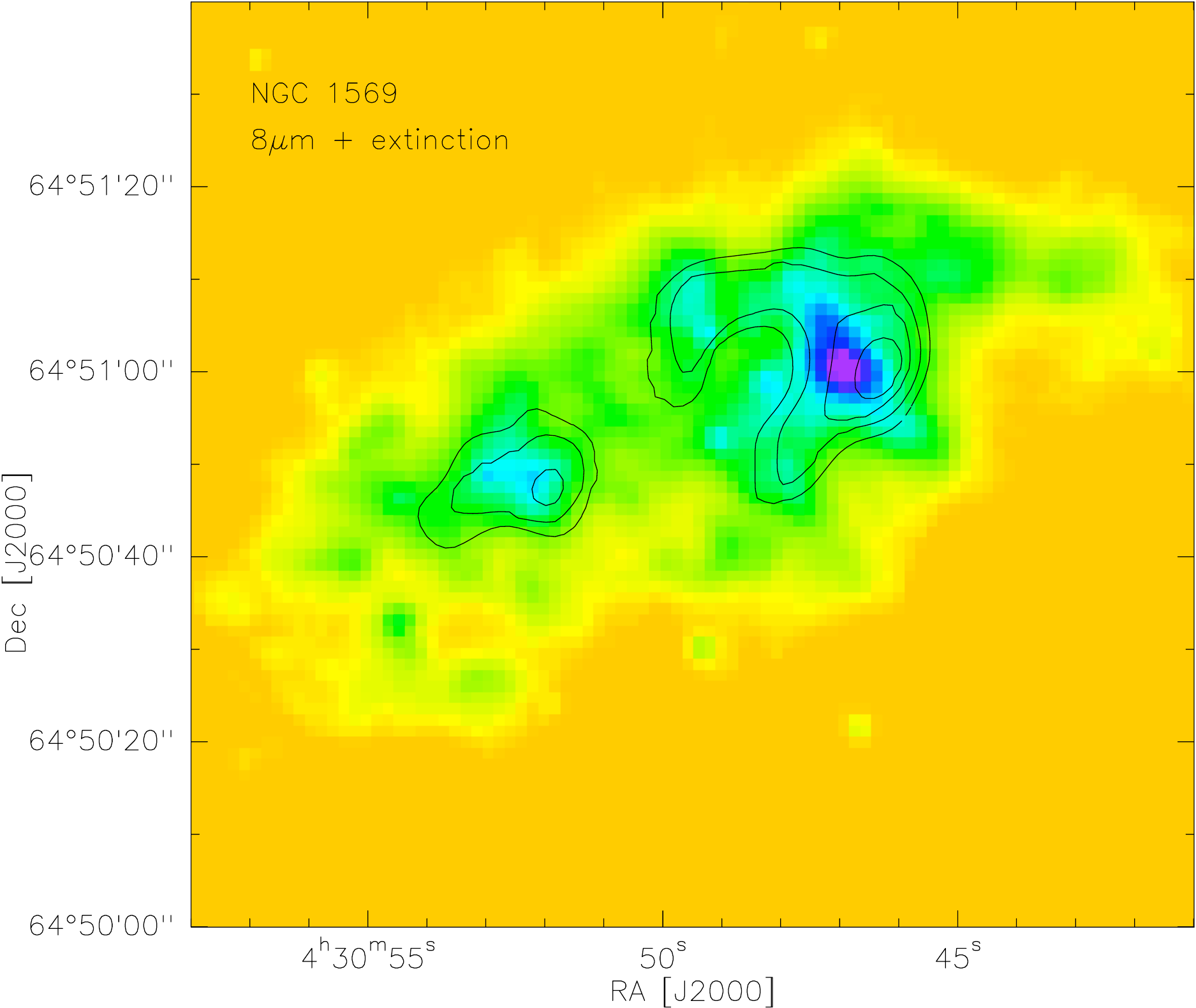} \quad
\includegraphics[width=6.cm,angle=270]{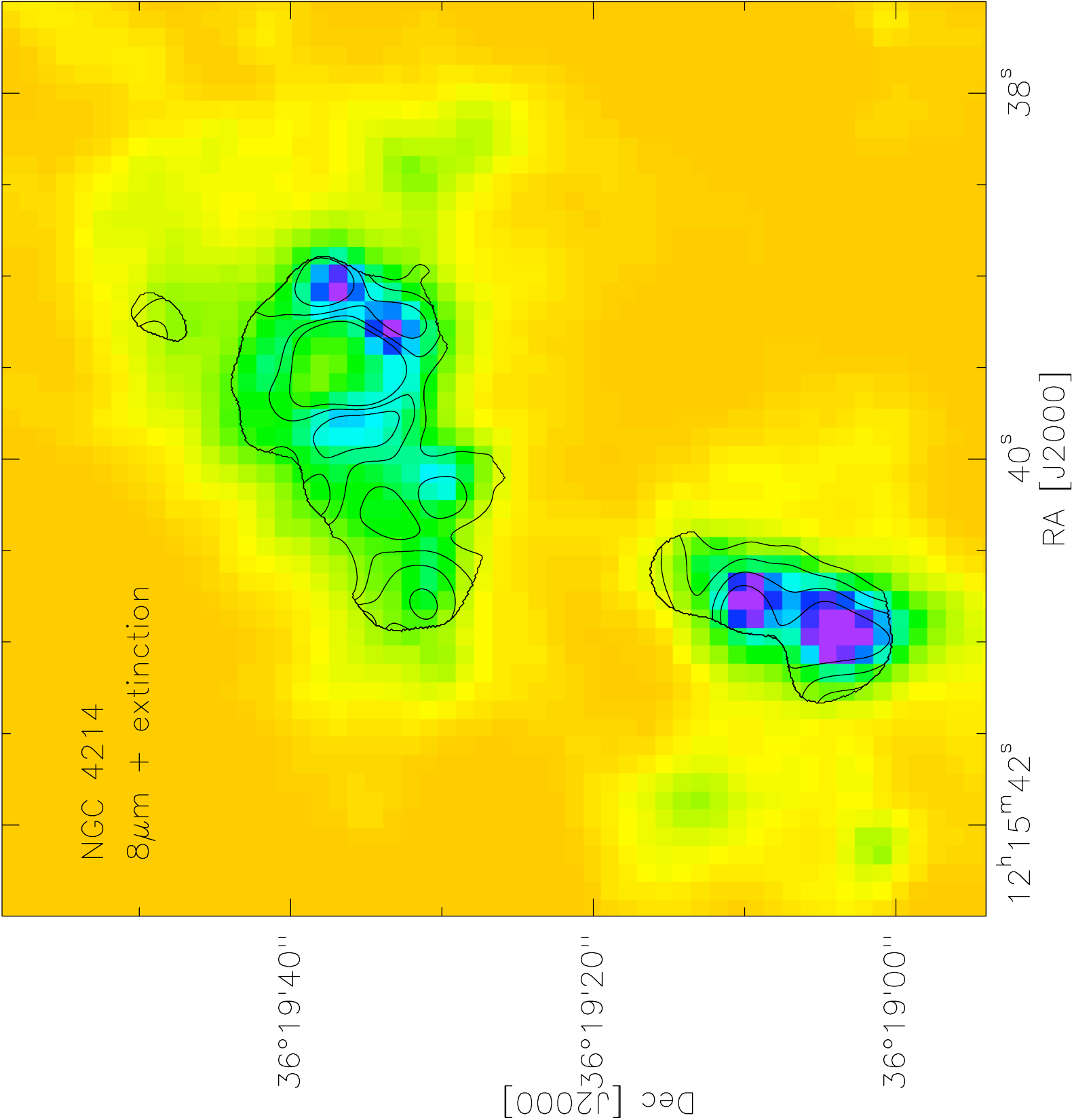} }
\centerline{\includegraphics[width=6cm, angle=270]{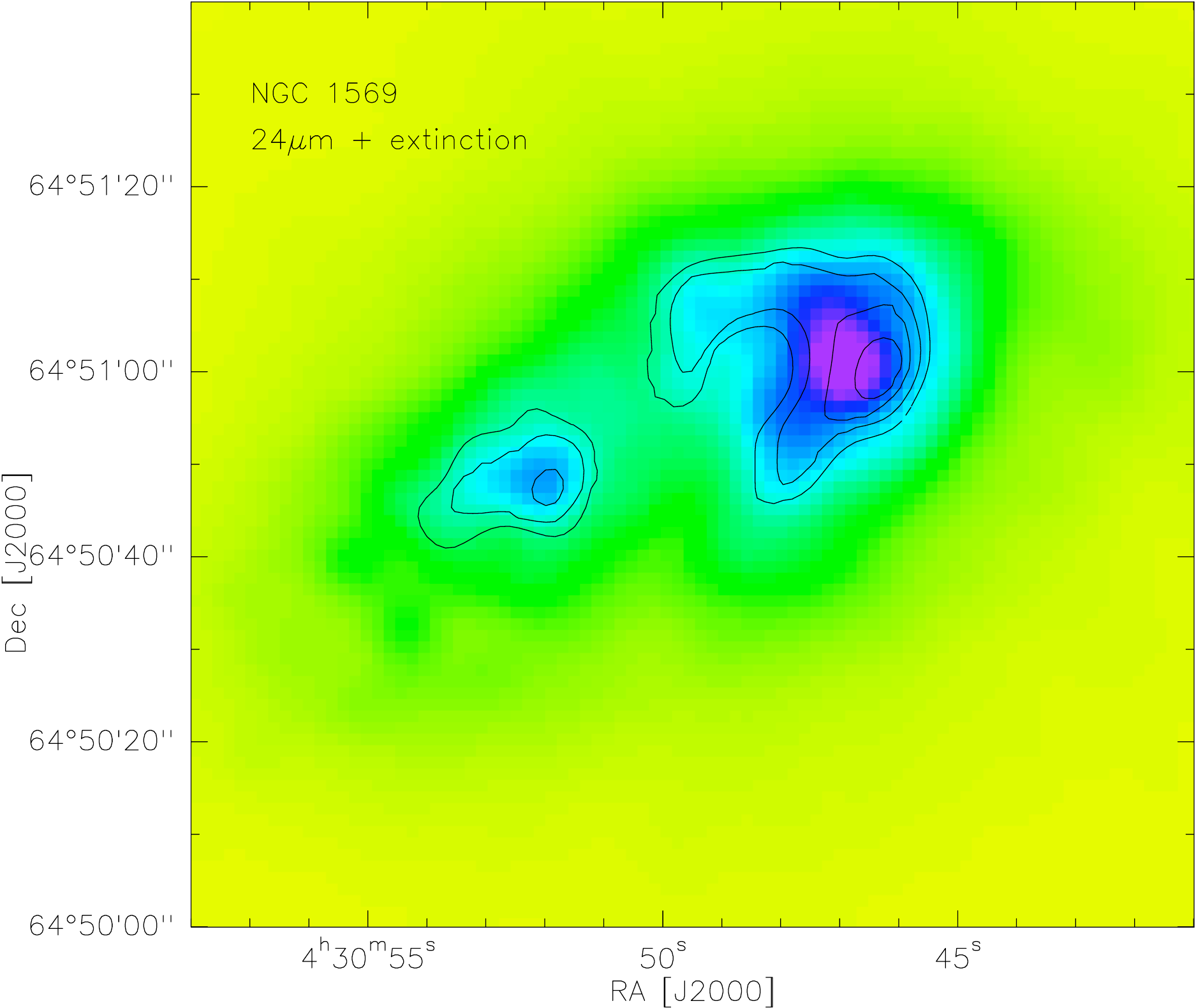} \quad
\includegraphics[width=6.cm,angle= 270]{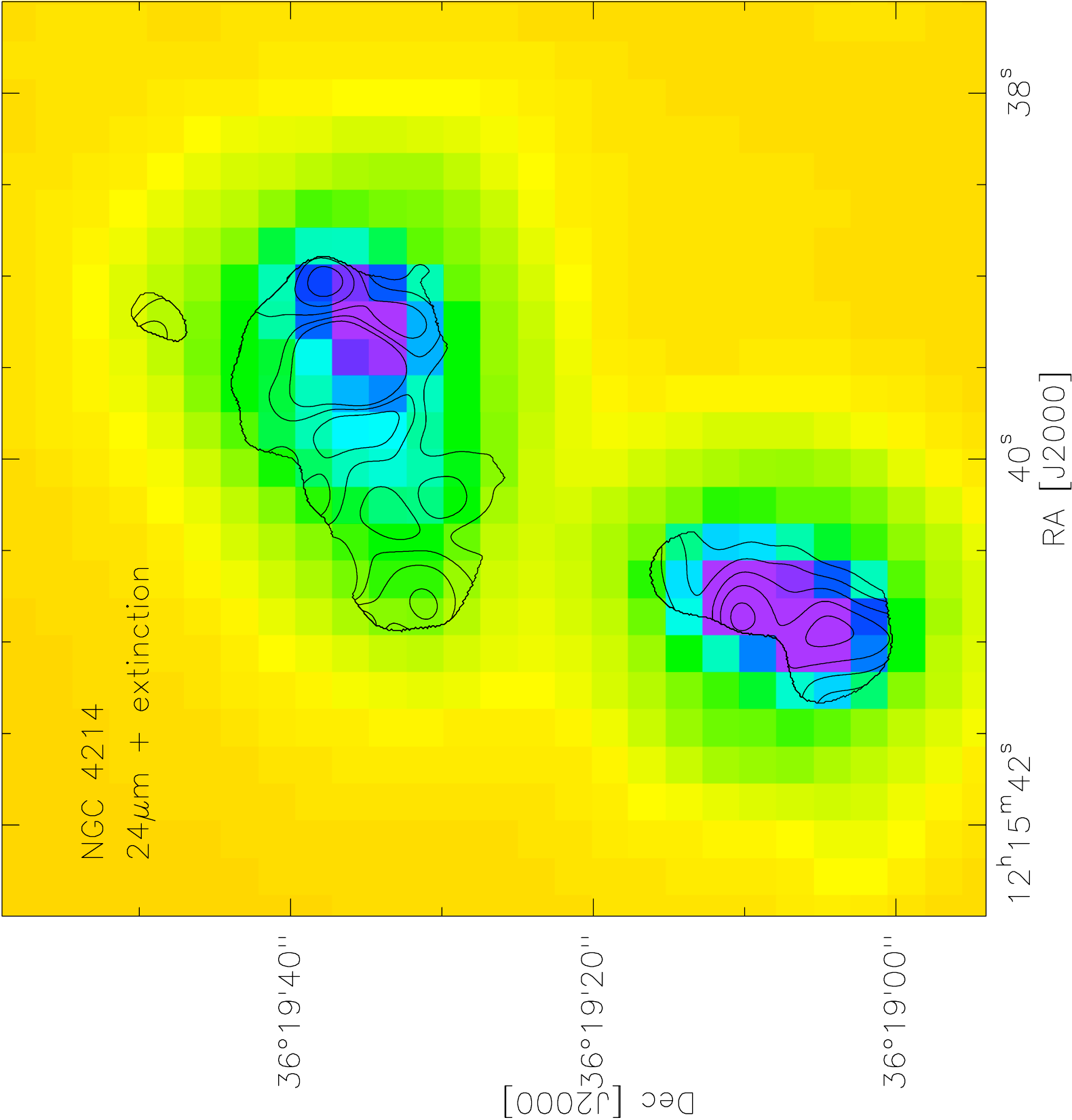} }
\centerline{ \includegraphics[width=6cm, angle=270]{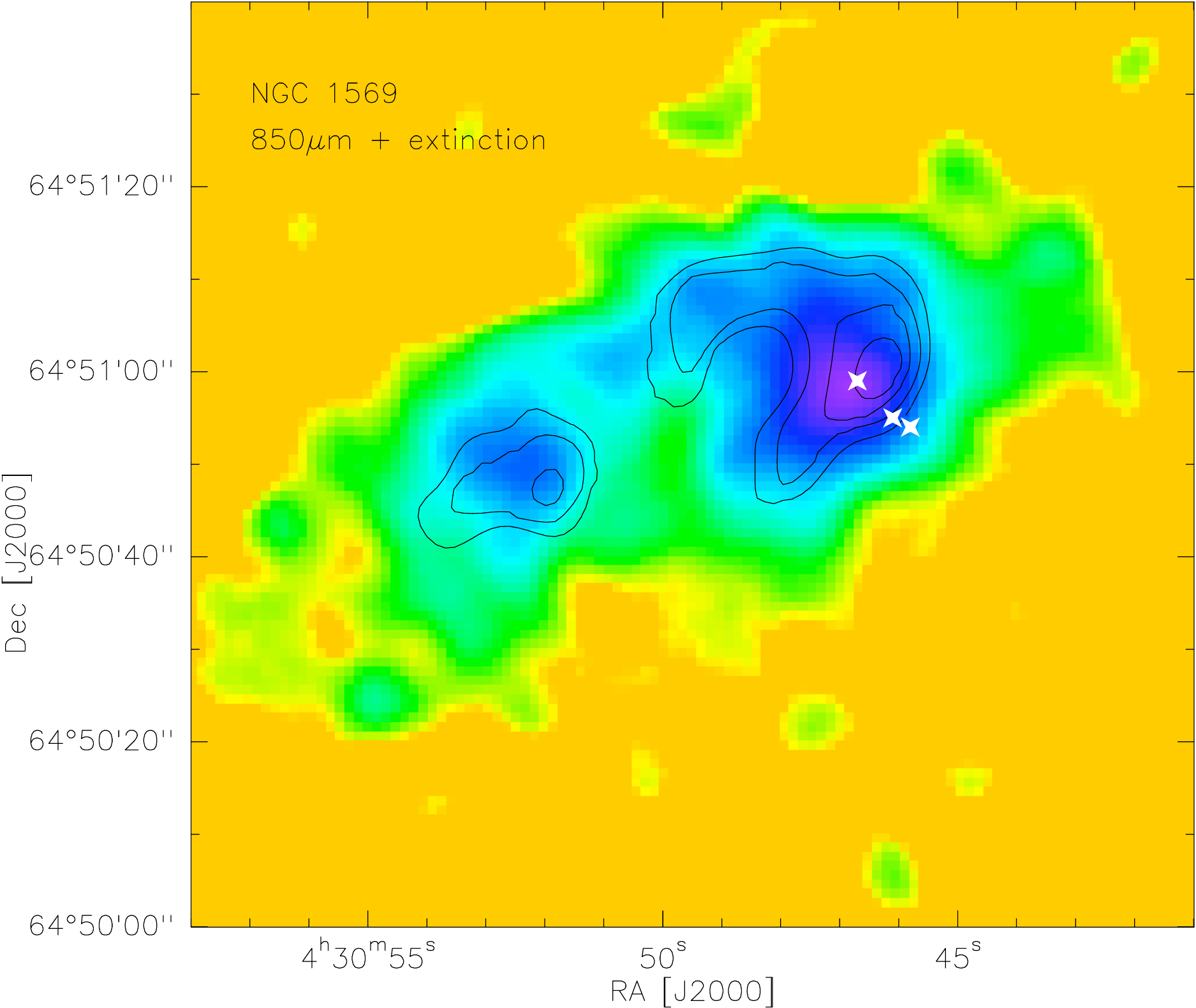} \quad
 \includegraphics[width=6.cm,angle= 270]{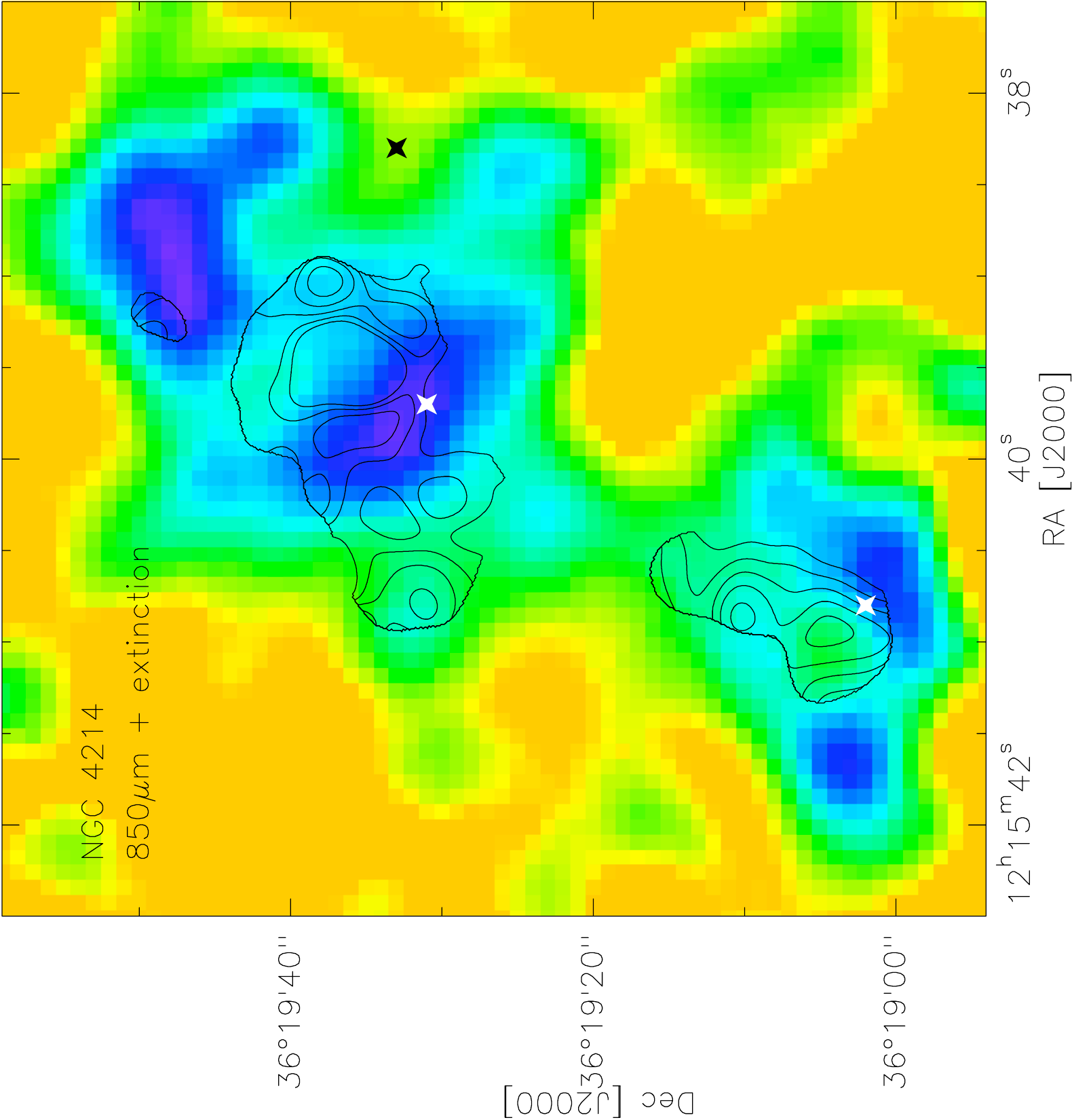}}
 \caption{Intrinsic extinction (contours levels at 0.1, 0.3, 0.5 and 0.7 mag for NGC 1569 and
 at 0.25, 0.45 and 0.65 mag for NGC 4214),  
 overlaid on the emission of  8, 24 and 850\mi\  (colour) for
 NGC 1569 ({\it left})  NGC 4214  ({\it right}).  
 The 8 and 24 \mi\ maps are from the {\it Spitzer} archive, and the 850 \mi\ maps are observed
 with SCUBA at the James Clerk Maxwell Telescope.
 The spatial resolution is 6'' (extinction and 24\mi),
1.4'' (8 \mi) and 15'' (850 \mi).
The crosses in the lowest panels  indicate the positions of the strongest 
peaks in the CO(1-0) distribution
(Taylor et al. 1999 for NGC 1569 and Walter et al. 2001 for NGC 4214).
 }
 
   \label{fig1}
\end{center}
\end{figure}







\begin{thebibliography}{}


\bibitem[]{Armus88} Armus, L., Heckman, T.M., Miley, G.K., 1989, ApJ, 347, 727

\bibitem[]{Calzetti05} Calzetti, D., Kennicutt, R.C., Bianchi., L., et al., 2005, ApJ, 633, 871

\bibitem[]{Calzetti07}  Calzetti, D., Kennicutt, R.C., Engelbracht, C.W., 2007, ApJ 666, 870

\bibitem[]{Caplan86} Caplan, J., Deharveng, L, 1986, A\&A,  155, 297

\bibitem[]{Israel88} Israel, F. P., A\&A, 194, 24

\bibitem[]{Kiuchi04} Kiuchi, G., Ohta, K., Sawicki, M., Allen, M., 2004, ApJ, 128, 2743

\bibitem[]{Kobulnicky97} Kobulnicky, H.A., Skillman, E., 1997, ApJ, 489, 636

\bibitem[]{Lisenfeld02} Lisenfeld, U., Israel, F.P., Stil, J.M., Sievers, A., 2002 A\&A, 382, 860

\bibitem[]{maiz02} Ma\'iz-Apell\'aniz, J., Cieza, L., MacKenty, J. W., 2002, AJ, 124, 1601

\bibitem[]{Martin98} Martin, C.L., 1998, ApJ, 506, 222

\bibitem[]{Perez06} P\'erez-Gonz\'alez, P. G., Kennicutt, R.C., Gordon, K.D., 2006, ApJ, 648, 987

\bibitem[]{Relano06}  Rela\~no, M., Lisenfeld, U., V\'ilchez, J.M., Battaner, E., 2006, A\&A, 452, 413

\bibitem[]{Relano07}  Rela\~no, M., Lisenfeld, U., P\'erez-Gonz\'alez, P. G., V\'ilchez, J.M., Battaner, E., 2006, A\&A 667, L141

\bibitem[]{schlegel98} Schlegel, D.J., Finkbeiner D. P., Davis, M., 1998, ApJ, 500, 525

\bibitem[]{walter01} Walter, F., Taylor, C. L., H\"uttemeister, S., Scoville, N., McIntyre, V., 
2001, ApJ, 121, 727

\bibitem[]{Zubkko04} Zubko, V., Dwek, E., Arendt, R.G., ApJS, 152, 211

\bibitem[]{Taylor99} Taylor, C.L., H\"uttemeister, S., KLein, U., Greve, A., 1999, A\&A, 349, 424

\end{thebibliography}
\end{document}